\documentclass[12pt]{article}

\usepackage{graphicx,amssymb,url,mathrsfs, amsmath}
\usepackage{eucal}
\usepackage{wrapfig}
\usepackage{boxedminipage}
\usepackage{setspace}
\usepackage{subfigure}
\def\a  {\alpha}                
       \def\d  {\delta}        \def\D  {\Delta}
\def\e  {\epsilon}        \def\k  {\kappa}
\def\l  {\lambda}             \def\m  {\mu}
\def\n  {\nu}          \def\s  {\sigma}        
\def\t  {\tau}                 
   \def\w  {\omega}

\newcommand{\cala}{\mbox{${\cal A}$}} 
 
 \newcommand{\calf}{\mbox{${\cal F}$}}

\def\IR{{\hbox{{\rm I}\kern-.2em\hbox{\rm R}}}}
\def\IB{{\hbox{{\rm I}\kern-.2em\hbox{\rm B}}}}
\def\IN{{\hbox{{\rm I}\kern-.2em\hbox{\rm N}}}}
\def\IC{\,\,{\hbox{{\rm I}\kern-.59em\hbox{\bf C}}}}
\def\IZ{{\hbox{{\rm Z}\kern-.4em\hbox{\rm Z}}}}
\def\IP{{\hbox{{\rm I}\kern-.2em\hbox{\rm P}}}}
\def\IH{{\hbox{{\rm I}\kern-.4em\hbox{\rm H}}}}
\def\ID{{\hbox{{\rm I}\kern-.2em\hbox{\rm D}}}}

\def\be{\begin{equation}}
\def\ee{\end{equation}}
\def\ba{\begin{eqnarray}}
\def\ea{\end{eqnarray}}

\def\half{\frac{1}{2}}

\newcommand{\inv}[1]{\frac{1}{#1}}

\newcommand{\ud}{\mbox{${\mathrm{d}}$}}

\def\dell{\partial}

\newcommand{\abs}[1]{\left| #1 \right|}

\def\dg{{\dagger}}
\def\Tr{{\rm tr}\,}

\def\nn{\nonumber}
\def\ea{{\it et al}. }

\newcommand{\Mkk}{M_{\rm KK}}
\newcommand{\wt}{\widetilde}
\newcommand{\wh}{\widehat}

\begin{document}

\begin{titlepage}

%\begin{flushright}
%  {\tt hep-th/0608046}
%\tt {FileName:FF1.tex} \\
% {\tt \today}
%\end{flushright}
\vspace{0.5in}

\begin{center}
{\large \bf The Nucleon as a Holographic Cheshire Cat}~\footnote{To appear in {\it Gerry Brown memorial volume}}\\
\vspace{10mm}
Ismail Zahed\\
\vspace{5mm}
{\it  Department of Physics and Astronomy, \\Stony Brook University, NY 11794}\\
\vspace{10mm}
  {\tt \today}
\end{center}
\begin{abstract}
The Cheshire cat principle emerges naturally from the holographic approach of the nucleon
in terms of a bulk instanton. The cat hides in the holographic direction. I briefly review the 
one-nucleon problem in the holographic limit.
\end{abstract}

\end{titlepage}

\renewcommand{\thefootnote}{\arabic{footnote}}
\setcounter{footnote}{0}

%\tableofcontents
%\newpage

%%%%%%%%%%%%%%%%%%%%%%%%%%%%%%%%%%%%%%%%%%%%%%%%%%%%%%%%%%%%%%%%%%%%%%%%%%%
%%%%%%%%%%%%%%%%%%%%%%%%%%%%%%%%%%%%%%%%%%%%%%%%%%%%%%%%%%%%%%%%%%%%%%%%%%%
%%%%%%%%%%%%%%%                 BODY                    %%%%%%%%%%%%%%%%%%%
%%%%%%%%%%%%%%%%%%%%%%%%%%%%%%%%%%%%%%%%%%%%%%%%%%%%%%%%%%%%%%%%%%%%%%%%%%%
%%%%%%%%%%%%%%%%%%%%%%%%%%%%%%%%%%%%%%%%%%%%%%%%%%%%%%%%%%%%%%%%%%%%%%%%%%%

\section{In memorium}
{\it This paper is dedicated to the memory of Gerry Brown who has been my mentor
and colleague for the past three decades. One day in the fall of 1982,
my advisor Michel Baranger informed me that he has invited
Gerry Brown to give a talk at MIT and meet me. He indicated briefly
that Gerry was an old friend of his,  that competed historically with him in 
the early days of the Lamb-shift calculation.  (Later I learned from Gerry that Baranger
beat him on the 10 Mega-cycle correction to the Feynman and Schwinger
Lamb-shift landmark calculation. However, Gerry was always very proud of
his higher-order corrections to the Lamb-shift in Coulombic atoms. He would
always go in great details and pride about how he carried this important
and difficult calculation, and also the physical effort it took him shuttling back-and-forth in Birmingham
between the computing facility and his office.)

I met Gerry in the fall of 1982, the morning after a tumultuous talk at the Laboratory of Nuclear Science
at MIT where Gerry was talking about the excited states of the little bag model. Needless
to say that little bags where not popular in the large bag sanctum that afternoon.
The meeting with Gerry took place in Feshbach office with Baranger and later Kerman present. 
I had a very good time discussing my several research projects with Gerry on the board.
We departed at lunch and Gerry wished me well in my work.

Just after the new year in 1983 and out of the blue,  I received a letter from Gerry offering
me a 3-year position in Stony Brook. I was stunned since I have not applied to any place,
let alone thought about graduating after just 3 years at MIT. After consulting with Baranger,
I wrote back to Gerry accepting his offer. Upon graduation in the Spring of 1983,  I was
also offered an NSF  NATO fellowship at the Niels Bohr Institute.  Gerry immediately indicated
that I should take it and his offer would still be valid at the end of the fellowship. He also
mentioned that he will be in Copenhagen in the fall of 1983 as part of his dual
appointment at  the Nordic Institute.

In the summer of 1983  I flew from Boston to Copenhaguen a country I never visited before.
I arrived at the Niels Bohr at around noon time. Upon checking at the secretarial office, I was
informed that Gerry has asked that I wait for him to be picked up.  Soon after the secretary entered the 
sounding Morse-like call code on the famed inter-phone at the Niels Bohr, Gerry showed up to
welcome me in this new place. He helped me carry my suitcase across the hall-ways of the institute
introducing me to luminaries of the place. We only stopped by Mottelson office for an official
introduction as I was going to be their NSF fellow for the next year.  After leaving my suitcase
in Gerry office we headed to the cafeteria where two tables were full of Gerry students, postdocs
and collaborators. I was stunned by the number of people around him.

After lunch Gerry asked me to follow him to his office.  There we started chatting about some physics, 
while in typical Gerry style he pulled out two papers from his brief-case and handed them to me. He
said that on his way to Copenhagen he passed by Princeton and there he met Witten and Nappi who
told him about their new work on the Skyrme model. The two papers were the by now famed work by
Witten and also Adkins, Nappi and Witten on the Skyrmion. Gerry asked me to explain them to him. 
While, I was about to be briefed on the logistics of the place Jim Lattimer and Gerry Cooperstein
came in. I told Gerry that I would be able to figure out the place and left.

The following many weeks I had the pleasure to discuss the papers with Gerry and his students
and collaborators. It was the beginning of a great journey by Gerry side.
 I owe much to Gerry in terms of mentoring, supporting and counseling during all my years
with him.  Gerry sense of physics and wit is unmached. More importantly his humanity as
measured by the amount of care and kindness he has shown to many of his students and 
collaborators is legend. For Gerry we were part of his family. For me he was one of the last eagles.}

\section{Introduction}

Historically, quark bag models were simplified models of hadrons
consisting of free quarks and gluons confined to a bag because of
asymptotic freedom, and dressed up by mostly pions to account
for the pion tail of baryons. The bag radius was initially considered 
measurable, with the current Jefferson facility being tasked to 
measure it. Two competing pictures emerged: the original MIT 
bag model with a large bag radius surrounded by a bare vacuum,
 and the Stony-Brook bag model with a small bag radius surrounded 
 by pions~\cite{BAGS}.

It turns out that this delineation is unphysical at low energy,
as demonstrated in the Cheshire cat principle~\cite{CAT}. Quantum effects and
anomalies cause most of the charges (fermionic, axial, etc.)
to leak making the bag boundary immaterial~\cite{LEAK}, much 
like the smile of the Cheshire cat in "Alice in wonderland"~\cite{ALICE}.
The Skyrme model is an example of this  principle whereby the
unphysical bag radius is reduced to zero size \cite{SKYRMION}. 

The Skyrme model has recently re-emerged from holographic QCD
in the double limit of large $N_c$ and strong coupling t'Hooft coupling $\l=g^2N_c$
~\cite{SAKAI,HRYY}. The Skyrmion is the line-integral of a flavored instanton 
 located along the holographic or 5th-dimension (see Eq.~\ref{HOLO}) below).
Baryon number at the boundary is dual to instanton 
number in bulk~\cite{WITTEN}. Although the holographic correspondence
is only proved for certain conformal gauge theories at the boundary
~\cite{MALDACENA}, it is usually assumed for non-conformal theories
such as QCD. Many of the arguments presented below have been worked out in 
~\cite{KZ,IZ}.

\begin{figure}[]
  \begin{center}
    \includegraphics[width=11cm]{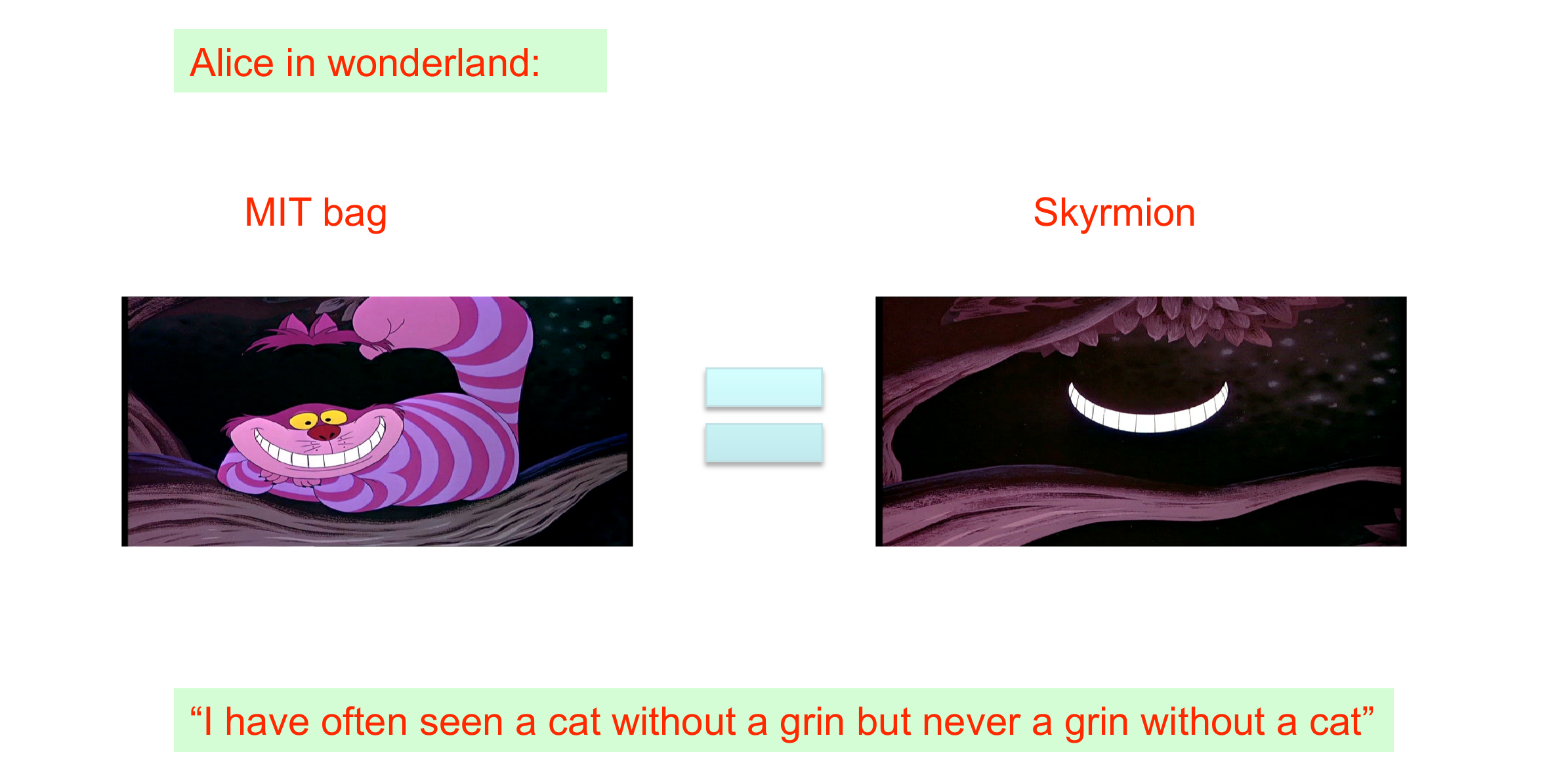}
  \caption{Alice's Cheshire Cat.}
  \label{Fig:fig1XX}
  \end{center}
\end{figure}
\begin{figure}[]
  \begin{center}
    \includegraphics[width=11cm]{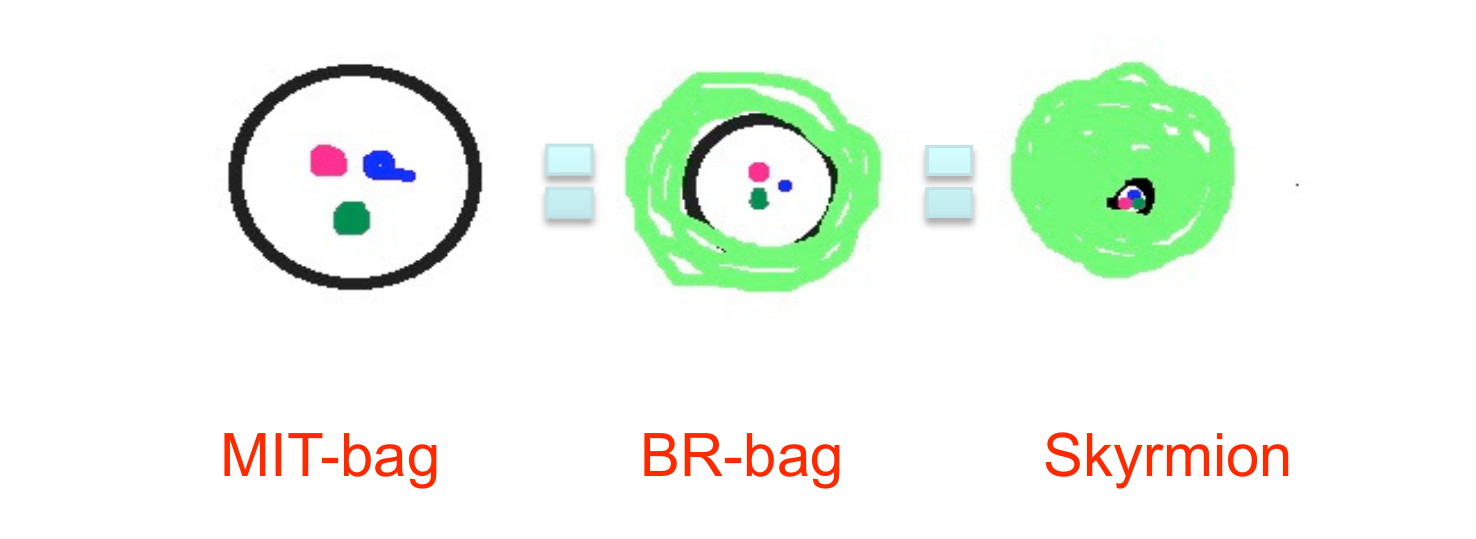}
  \caption{The Cheshire Cat Principle.}
  \label{Fig:fig1X}
  \end{center}
\end{figure}

\section{The Skyrmion from the Instanton}

In the large $N_c$ limit baryons are chiral solitons. A firts principle framework for 
discussing chiral solitons at large ${}^\prime$t Hooft coupling $\lambda=g^2N_c$ is the use
of the probe $D8$-$\overline{D8}$ flavor branes in the $D4$ color brane set up as 
discussed by Sakai and Sugimoto~\cite{SAKAI}. After reduction, the result is an 
effective flavor Yang-Mills-Chern-Simons theory in a 5 dimensional curved 
background. More specifically, the leading contributions  in a $1/\l$ expansion 
of the D-brane  Born-Infeld (DBI) effective action on D8 is~\cite{SAKAI},
\begin{eqnarray}
\label{CS}
&&S = S_{YM} + S_{CS}\ ,  \label{YM-CS}\\
&&S_{YM} = - \k \int d^4x dZ \ \Tr \left[\half K^{-1/3}
\calf_{\m\n}^2 + \Mkk^2 K
\calf_{\m Z}^2 \right] \ , \label{YM}\nonumber \\
&&S_{CS} = \frac{N_c}{24\pi^2}\int_{M^4 \times R}
\w_5^{U(N_f)}(\cala) \ ,\nonumber 
%\label{REDUCED}
\end{eqnarray}
where $\m,\n = 0,1,2,3$ are 4D indices and the 5th  coordinate is $Z$. 
Also $\k=\lambda N_c/216\pi^3$. 
The confining scale in this gravity dual reduction is the Kaluza-Klein 
compactification of the $D4$ branes or $M_{KK}$. Throughout, all units
will be expressed in terms of $M_{KK}=1$ for convenience. The effect
of the 5-dimensional gravity due to the $D4$ branes in bulk is the warping 
$K=1+Z^2$. $\cala$ is the 5D $U(N_f)$ 1-form gauge field and $\calf_{\m\n}$
and $\calf_{\m Z} $ are the components of the 2-form field
strength $\calf = \ud \cala -i \cala \wedge \cala$.
$\w_5^{U(N_f)}(\cala)$ is the Chern-Simons 5-form for the $U(N_f)$
gauge field
\begin{eqnarray}
  \w_5^{U(N_f)}(\cala) = \Tr \left( \cala \calf^2 + \frac{i}{2} \cala^3 \calf - \inv{10} \cala^5
  \right)\ ,
\end{eqnarray}

In (\ref{CS}) baryons are flavor SU(2) instantons of the reduced 
Yang-Mills-Chern-Simons theory. At large $\l=g^2N_c$, the instanton is 
forced by the gravitational warping or $K$ to localize near $Z\approx 0$
for otherwise the mass term in (\ref{CS}) is large. On the other
hand, the topological Chern-Simons term is repulsive preventing 
the instanton to shrink to zero size. The balance of these two effects
is an instanton of size  $Z\approx 1/\sqrt{\l}\ll 1$ in units of $M_{KK}$~\cite{SAKAI}.

For small size instantons localized near $Z\approx 0$,
the effects of the gravitational warping on the instanton
can be neglected to leading order. Thus the instanton SU(2) flavor configuration
${\mathbb{A}}_M$ and its supporting U(1) Coulomb potential
$\widehat{\mathbb{A}}_M$ read

\begin{eqnarray}
\widehat{\mathbb{A}}_0 = -\frac{1}{8 \pi^2 a \l}\frac{2\rho^2 +
\xi^2}{(\rho^2 + \xi^2)^2} \ , \qquad\qquad\qquad
\mathbb{A}_M = \eta_{iMN}\frac{\sigma_i}{2}\frac{2
x_N}{\xi^2 + \rho^2} \ ,   
\label{ADHM2}
\end{eqnarray}
with all other gauge components set to zero. Here $\xi^2=\vec{x}^2+Z^2$ and $\rho\sim 1/\sqrt{\l}$
is the instanton size. We refer to~\cite{SAKAI} for more details on the relevance 
of this configuration for baryons. The configuration has  spherical 
symmetry and satisfies
\begin{eqnarray}
(\mathbb{R}\mathbb{A})_Z = \mathbb{A}_Z(\mathbb{R} \vec{x}) \ ,
\qquad (\mathbb{R}^{ab}\mathbb{A}^{b})_i =
\mathbb{R}^T_{ij}\mathbb{A}_j^a(\mathbb{R}\vec{x})\,\,,
\end{eqnarray}
with $\mathbb{R}^{ab}\tau^b=\Lambda^+\tau^a\Lambda$ a rigid SO(3) rotation,
and $\Lambda$ is its SU(2) version. 
The classical baryon  is the projected map or holonomy along the Z-direction

\begin{eqnarray}
U^{\mathbb{R}}(x)=\Lambda{\bf P}{\rm exp}
\left(-i\int_{-\infty}^{+\infty}dZ\,\mathbb{A}_Z\right)\Lambda^+\,\,.
\label{HOLO}
\end{eqnarray}
The corresponding Skyrmion is
$U^{\mathbb{R}=1}(\vec{x})=e^{i\vec{\tau}\cdot\vec{x}{\bf F} (\vec{x})}$ with
the profile ${\bf F} (\vec{x})={\pi |\vec{x}|}/{\sqrt{{\vec{x}}^2+\rho^2}}$.

The quantum baryon emerges from a semiclassical quantization of the 
classical source (\ref{ADHM2}-\ref{HOLO}). To achieve this, we define

\begin{eqnarray}
A_M(t,x,Z) = \mathbb{R}(t) \left[\mathbb{A}_M(x-X_0(t),Z-Z_0(t))
+ C_M(t,x-X_0(t),Z-Z_0(t))\right] \ , \label{SEM1}
\end{eqnarray}
The collective coordinates 
$\mathbb{R}, X_0, Z_0$ and the fluctuations $C$ in 
(\ref{SEM1}) form a redundant set. The redundancy is lifted 
by constraining the fluctuations to be orthogonal to the
zero modes. This can be achieved either rigidly~\cite{ADAMI} 
or non-rigidly~\cite{VERSHELDE}. We choose the latter 
via the constraint $\int_{x=Z=0} d{\hat{\xi}} C\,G^B\mathbb{A}_M$
with $(G^B)^{ab}=\epsilon^{aBb}$ the real generators of $\mathbb{R}$.

\begin{figure}[]
  \begin{center}
    \includegraphics[width=11cm]{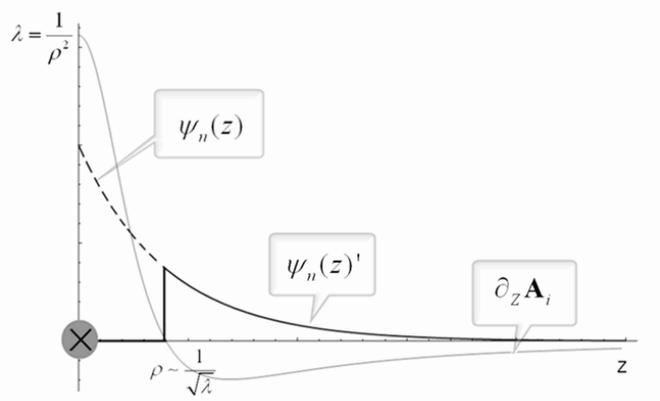}
  \caption{A vector fluctuation $\psi_n(Z)$ versus a quasi-zero mode $\dell_Z \mathbb{A}_i$.
  In the non-rigid gauge the new vector fluctuation is denoted by $\psi_n^\prime(Z)$. See text.}
  \label{Fig:fig1}
  \end{center}
\end{figure}

For the quasi-zero mode in $Z$, the non-rigid constraint is more natural to
implement since this mode is mostly localized near $Z\approx 0$.  It is also local
and thus preserves causality. The vector fluctuations at the origin linearize through the usual modes
\begin{eqnarray}
d^2\psi_n/dZ^2= -\l_n\psi_n \ , \label{SEM6}
\end{eqnarray}
with $\psi_n(Z)\sim e^{-i\sqrt{\l_n}Z}$. In the spin-isospin 1
channel they are easily confused with 
the zero mode: $\partial_Z\mathbb{A}_i$
near the origin as we show in Fig.~\ref{Fig:fig1}.
Using the non-rigid constraint,  we define a new set of modes
\begin{eqnarray}
\psi_n'(Z)=\theta(|Z|-Z_C)\psi_n(Z) \ , \label{SEM7}
\end{eqnarray}
with $Z_C\sim\rho\sim 1/\sqrt{\l}$ which is the origin for
large $\l$. The  baryon at small $\xi<|Z_C|$ is described 
by a flat or uncurved instanton located at the origin (Cheshire cat).
At large $\xi>|Z_C|$, the  instanton sources the vector meson
fields (cloud). This is a holographic realization of the Cheshire cat
principle~\cite{CAT} whereby $Z_C$ plays the role of the Cheshire
cat smile. In a way Alice$^\prime$s Cheshire cat hides in the holographic direction.

\section{The Baryonic Current}

To extract the baryon current or nucleon form factor,
we source the reduced 
action with $\hat{\cal V}_\mu(x)$ a $U(1)_V$ flavor 
field on the boundary in the presence of the vector
fluctuations or $C=\hat{v}$. The tree level baryonic 
current reads~\cite{KZ}

\begin{eqnarray}
J^\m_{B}(x) &=& -  \k K
\widehat{\mathbb{F}}^{Z \m} (x,Z)\,
(1-\sum_{n=1}^\infty\a_{v^n}\psi_{2n-1})
\Big|_{Z=B}  \nn \\
&&
 -\sum_{n,m}  \  m_{{v}^n}^2 a_{{v}^n} \psi_{2m-1}
\int d^4y\
 \k K  \widehat{\mathbb{F}}_{Z \n}(y,Z) \Delta^{\n \m}_{mn}(y-x) \Big|_{Z=B}
 \ .
  \label{Bcurrent2}
\end{eqnarray}
The massive vector meson propagator in Lorentz gauge is
\begin{eqnarray}
\Delta_{\m\n}^{mn}(x) = \int \frac{d^4 p}{(2\pi)^4} e^{-ipx}
\left[ \frac{- g_{\m\n} - p_\m p_\n / m_{v^n}^2}{p^2 + m_{v^n}^2}
\d^{mn}\right] \ ,
\end{eqnarray}
The first contribution in (\ref{Bcurrent2}) is the nucleon core instanton.
The second contribution sums up the cloud of all vector mesons. The
chief observation is that the core coupling vanishes due to the exact sum
rule
\begin{eqnarray}
\sum_{n=1}^{\infty} \a_{v^n}\psi_{2n-1} = 1 \ , \label{sumrule1}
\end{eqnarray}
following from closure in curved space. {\it No cat is left}:
vector Meson Dominance (VMD) is exact
in holography. A similar argument holds for the pion electromagnetic 
form factor ~\cite{SAKAI}. The results presented in this section were derived in~\cite{KZ}.
They were independently arrived
at in~\cite{HASHI} using the strong coupling source quantization.
They also support the bottom up effective approach 
described in~\cite{RHO} using the heavy nucleon expansion.

If we were to truncate the resonance contributions to the lowest vector mesons,
the core contribution is non-zero. In this case, the deleniation of the Cheshire cat
smile is no longer arbitrary. A specific position of the smile means a minimal number
 of vector mesons in the truncated cloud. 
 For many years Gerry Brown has been advocating the 50/50 scenario for the baryon
form factor using both phenomenology and his democratic principle.

\begin{figure}[]
  \begin{center}
    \includegraphics[width=11cm]{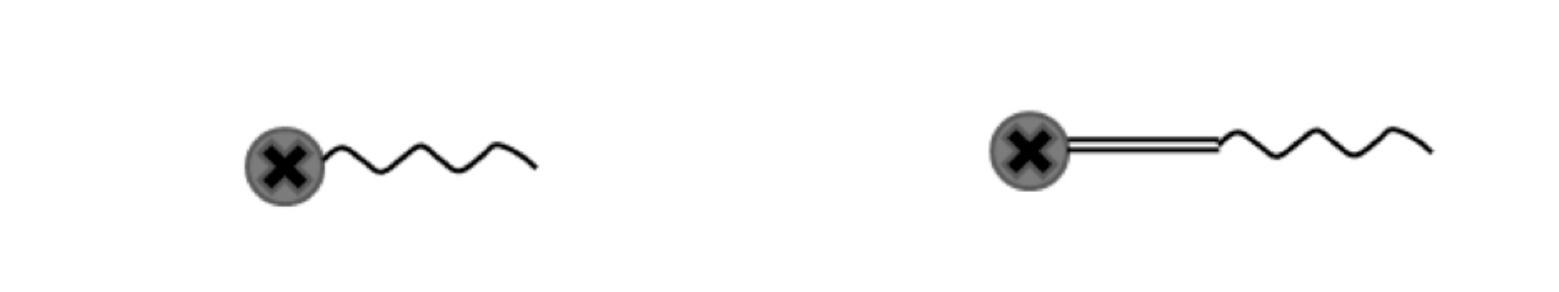}
  \caption{Gerry Brown 50/50 scenario:  Direct (left)  + VMD (right). See text.}
  \label{Fig:FF}
  \end{center}
\end{figure}

\section{Acknowledgments}
I thank my collaborator K.Y. Kim  for many discussions
on these issues. This work was supported in part by US-DOE grants 
DE-FG02-88ER40388 and DE-FG03-97ER4014.

\end{document}